\documentclass[prl,aps,twocolumn]{revtex4}
\usepackage{amsfonts}
\usepackage{amssymb}
\usepackage{graphicx}
\usepackage{mathrsfs}
\usepackage{array}
\usepackage{amsmath}
\usepackage{lscape}
\usepackage{bbold}
\usepackage{ucs}
\usepackage{color}

\begin{document}

\title{Emergence of Compositional Representations in Restricted Boltzmann Machines}

\author{J. Tubiana, R. Monasson}

\affiliation{Laboratoire de Physique Th\'eorique, Ecole Normale Sup\'erieure and CNRS, PSL Research, Sorbonne Universit\'es UPMC, 24 rue Lhomond,75005 Paris, France}

\date{\today}

\begin{abstract}
Extracting automatically the complex set of features composing real high-dimensional data is crucial for achieving high performance in machine--learning tasks. Restricted Boltzmann Machines (RBM) are empirically known to be efficient for this purpose, and to be able to generate distributed and graded representations of the data.  We characterize the structural conditions (sparsity of the weights, low effective temperature, nonlinearities in the activation functions of hidden units,  and adaptation of fields maintaining the activity in the visible layer) allowing RBM to operate in such a compositional phase. Evidence is provided by the replica analysis of an adequate statistical ensemble of random RBMs and by RBM trained on the handwritten digits dataset MNIST.
\end{abstract}
\maketitle

\begin{figure}[t]
\begin{center}
\includegraphics[width=\linewidth]{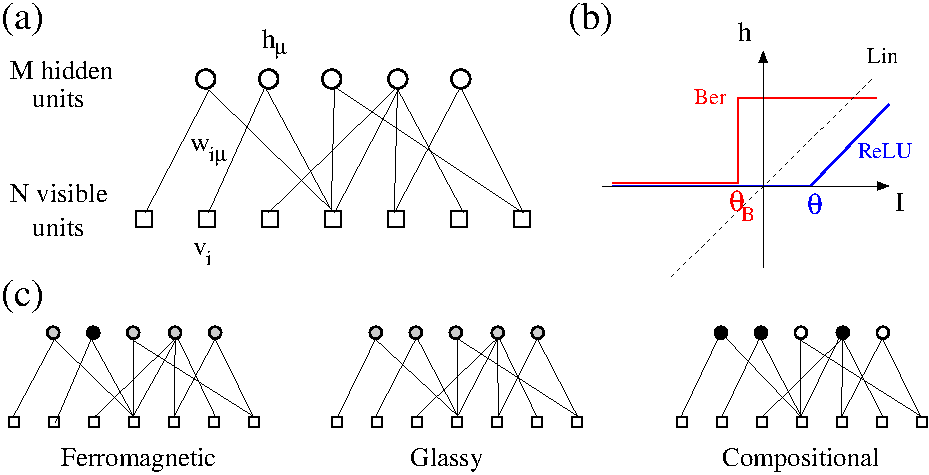}
\end{center}
\caption{{\bf (a)} Architecture of RBM. Visible ($v_i$, $i=1,\dots ,N$) and hidden ($h_\mu$, $\mu=1,..,M$) units are connected through weights ($w_{i\mu}$). {\bf (b)} Activation functions $\Phi$ of Bernoulli, Linear and Rectified Linear Units. The corresponding potentials are ${\cal U}^{Lin}(h)=\frac{h^2}2$; ${\cal U}^{Ber}(h)=h\, \theta_B$ if $h=0$ or $1$, and $+\infty$ otherwise;  ${\cal U}^{ReLU}(h)=\frac{h^2}2 + h\, \theta$ for $h\ge 0$, $+\infty$ for $h<0$. {\bf (c)} The three regimes of operation, see text. Black, grey and white hidden units symbolize, respectively, strong, weak and null activations.}
\label{fig:archi}
\end{figure}

Recent years have witnessed major progresses in supervised machine learning, {\em e.g.} in video, audio, image processing,... 
\cite{deeplearning}. Despite those impressive successes, unsupervised learning, in which the structure of data is learned without {\em a priori} knowledge still presents formidable challenges. A fundamental question is how to learn probability distributions that fit well complex data manifolds in high-dimensional spaces \cite{representationlearning}. Once learnt, such {\em generative models} can be used for denoising, completion, artificial data generation,... Hereafter we focus on one important generative model, Restricted Boltzmann Machines (RBM) \cite{smolensky,hinton}. In its simplest formulation a RBM is a Boltzmann machine on a bipartite graph, see Fig.~\ref{fig:archi}(a), with a visible (v) layer that represents the data, connected to a hidden (h) layer meant to extract and explain their statistical features. The marginal distribution over the visible layer is fitted to the data through approximate likelihood maximization \cite{training1,training2, training3, training4}.  Once the parameters are trained each hidden unit becomes selectively activated by a specific data feature; owe to the bidirectionality of connections  the probability to generate configurations of the visible layer where this feature is present is, in turn, increased. Multiple combinations of numbers of features, with varying degrees of activation of the corresponding hidden units allow for efficient generation of a large variety of new data samples.  However, the existence of such `compositional' encoding seems to depend on the values of the RBM parameters, such as the size of the hidden layer \cite{igel}. Characterizing the conditions under which RBM can operate in this compositional regime is the purpose of the present work.

In the RBM shown in Fig.~\ref{fig:archi}(a) the visible layer includes $N$  units $v_i$, with $i=1,\ldots,N$, chosen here to be binary ($=0,1$). Visible units are connected to $M$ hidden units $h_\mu$, through the weights $\{w_{i\mu}\}$. The energy of a configuration ${\bf v}=\{v_i\}, {\bf h} =\{h_\mu\}$ is defined through
\begin{equation}\label{erbm}
E[{\bf v}, {\bf h}]= -\sum_{i=1}^N \sum_{\mu=1}^M w_{i\mu}\,v_i\,h_\mu - \sum_{i=1}^N g_i\,v_i + \sum_{\mu=1}^M {\cal U}_\mu(h_\mu) \ ,
\end{equation}
where ${\cal U}_\mu$ is a potential acting on hidden unit $\mu$; due to the binary nature of the visible units their potential is fully characterized by a local field, $g_i$ in (\ref{erbm}). Configurations are then sampled from the Gibbs equilibrium distribution associated to $E$, $P[{\bf v},{\bf h}] =\exp(-E[{\bf v},{\bf h}])/Z$, where $Z$ is the partition function \cite{smolensky}. 

Given a visible configuration ${\bf v}$ the most likely value $h_\mu$ of hidden unity $\mu$ is a function of its input $I_\mu=\sum_{i=1}^N  w_{i\mu}\,v_i$: $h_\mu =\Phi_\mu(I_\mu)$, where the activation function $\Phi_\mu=({\cal U}'_\mu)^{-1}$ as can be seen from the minimization of $E$. Examples of $\Phi$ are shown in Fig.~\ref{fig:archi}(b). When $\Phi$ is linear, {\em i.e.} for quadratic potential ${\cal U}$ the probability $P[{\bf v},{\bf h}]$ is Gaussian in the hidden units, and the marginal distribution $P[{\bf v}]$ of the visible configurations ${\bf v}$ can be exactly computed \cite{nul}. It coincides with the equilibrium distribution of a Boltzmann machine with a pairwise interaction matrix $J_{ij}=\sum_\mu w_{i\mu}w_{j\mu}$, or, equivalently, of a Hopfield model \cite{hopfield}, whose $M$ patterns ${\bf w}_\mu$ are the columns of the weight matrix $\{w_{i\mu}\}$. 

Activation functions $\Phi$ empirically known in machine--learning literature to provide good results are, however, nonlinear.  Nonlinear $\Phi$ produce effective Boltzmann machines with high order ($>2$) multibody interactions between the visible units $v_i$. Two examples are shown in Fig.~\ref{fig:archi}(b): Bernoulli units, which take discrete 0,1 values, and Rectified Linear Units (ReLU) \cite{deeplearning}. Unlike Bernoulli units ReLU preserve information about the magnitudes of their inputs above threshold \cite{relu}; this property is expected for real neurons and ReLU were first introduced in the context of theoretical neuroscience \cite{treves}. 


We first report results from a training experiment of RBM with ReLU on the handwritten digits dataset MNIST \cite{MNIST}. Our goal is not to classify digits from 0 to 9, but to learn a generative model of digits from examples. Details about learning can be found in Supplemental Material, Section~I. Figure 2(a) shows typical `features' ${\bf w}_\mu=\{w_{i\mu}\}$ after learning. Each feature includes negative and positive weights, and is localized around small portions of the visible layer. These features look like elementary strokes, which are combined by the RBM to generate random digits (Fig.~2(b)). In each generated handwritten digit image $\hat S\simeq 240$ hidden units are silent ($h_\mu=0$), see histogram in Fig.~2(c). The remaining hidden units have largely varying activations, some weak and few very strong; we estimate the number of strongly activated ones through the participation ratio $\hat L=  [(\sum_\mu h_\mu^a)^2/\sum_\mu h_{\mu}^{2a}]$, with exponent $a=3$ as explained below. On average $\hat L\simeq  20$ elementary strokes compose a generated digit, see Fig.~2(c). Different combinations of strokes correspond to different variants of the same digits. Many of those variants are not contained in the training set, and closely match digits in the test set (Supplemental Material, Fig.~1(b)), hence showing the generative power of RBM.

\begin{figure}[t]
\begin{center}
\includegraphics[width=\linewidth]{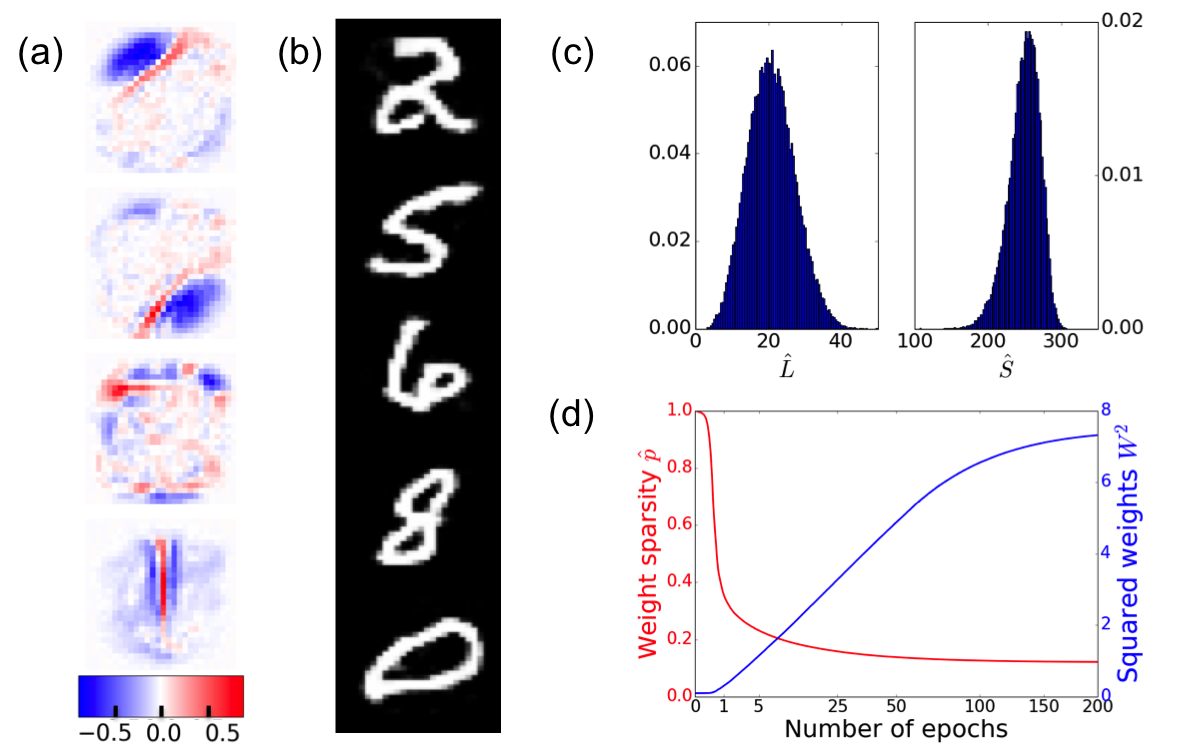}
\end{center}
\caption{Training of RBM on MNIST, with $N=28\times 28$ visible units and $M=400$ hidden ReLU.  {\bf (a)} Set of weights ${\bf w}_{\mu}$ attached to  four  hidden units $\mu$. {\bf (b)}  Averages of $\bf v$ conditioned to five hidden-unit configurations $\bf h$ sampled from the RBM at equilibrium. Black and white pixels correspond respectively to averages equal to $0$ and 1; few intermediary values, indicated by grey levels, can be seen on the edges of digits.  {\bf (c)}  Distributions of $\hat L $ (left) and $\hat S$ (right) in hidden-unit configurations at equilibrium. {\bf (d)} Evolution of the weight sparsity $\hat p$  (red) and the squared weight value $W_2$ (blue); The training time is measured in epochs (number of passes over the data set), and represented on a square--root scale. }

\label{fig:mnist}
\end{figure}

Learning is accompanied  by structural changes in RBM, which we track with two parameters: $\hat p = \frac{1}{MN} \sum_\mu  [(\sum_i w_{i\mu}^2)^2/\sum_i w_{i\mu}^4]$ and $W_2 = \frac 1M \sum_{i,\mu} w_{i\mu}^2$. Those parameters are proxies for, respectively, the fraction of nonzero weights and the effective inverse temperature, see  Supplemental Material, Section III. Figure 2(d) shows that $\hat p$ diminishes to small values $\sim 0.1$, whereas $W_2$ increases. While most weights become very small and negligible the remaining ones get large, in agreement with Fig.~2(a). Notice that sparsity is not imposed to obtain a specific class of features, {\em e.g.} as in \cite{olh96}, but naturally emerges through likelihood maximization across training. 
The presence of large weights implies that flipping visible units is associated to large energy costs. Visible units are effectively at very low temperature, as can be seen from the quasi binary nature of conditional averages in Fig.~2(b) and Supplemental Material, Fig.~5.

We argue below that these structural changes are not specific to MNIST-trained RBM but are generically needed to bring RBM towards a compositional phase, in which visible configurations are composed from combinations of a large number $L$ (typically, $1 \ll L \ll M$, as in Fig.~2(c))  of features encoded by simultaneously, strongly activated hidden units. Our claim is supported by a detailed analysis of a Random RBM (R-RBM) ensemble, in which the weights $w_{i\mu}$ are quenched random variables, with controlled sparsity and strength, and the magnitude of the visible fields and the values of the ReLU thresholds can be chosen. For adequate choices of these control parameters the compositional phase is thermodynamically favoured with respect to the ferromagnetic phase of the Hopfield model, where one pattern is activated \cite{ags}, and to the spin-glass phase, in which all hidden units are weakly and incoherently activated (Fig.~\ref{fig:archi}(c)).


In the R-RBM ensemble weights $w_{i\mu}$  are independent random variables, equal to $-\frac 1{\sqrt N},0,+\frac1{\sqrt N}$ with probabilities equal to, respectively, $\frac{p_i}2,1-p_i, \frac{p_i}2$; $p_i\in [0;1]$ sets the degree of sparsity of the weights attached to the visible unit $v_i$, high sparsities corresponding to small $p_i$.  The estimator $\hat p$ defined above (Fig.~2(d)) measures the fraction of nonzero weights, $p = \sum_i p_i/N$. This distribution was previously introduced to study parallel storage of multiple sparse items in the Hopfield model \cite{barra,barra2}. For simplicity the fields on visible units and the potentials acting on hidden units are chosen to be uniform, $g_i=g$ and ${\cal U}_\mu={\cal U}^{ReLU}$ (Fig.~1(a)). We define the ratio of the numbers of hidden and visible units, $\alpha=M/N$. 

Given a visible layer configuration $\bf v$, hidden units $\mu$ coding for features ${\bf w}_\mu$ present in $\bf v$ will be strongly activated: their inputs $I_\mu= {\bf w}_\mu\cdot {\bf v}$ will be strong and positive, comparable to the product of the norms of ${\bf w}_{\mu}$ ($\simeq \sqrt p$ for large $N$) and $\bf v$ (of the order of $\sqrt {p\,N}$), and, hence, will scale as $m\sqrt N$, where $m$, called magnetization, is finite. Most hidden units $\mu'$ have, however, features ${\bf w}_{\mu'}$ essentially orthogonal to $\bf v$, and receive inputs $I_{\mu'}$ fluctuating around 0, with finite variances. These scalings ensure that $\hat L$ defined above (Fig.~2(c)) will coincide with the number $L$ of strongly activated units when $N\to\infty$; choosing exponent $a=2$ in $\hat L$ rather than $a=3$ would have introduced biases coming from weakly activated units (Supplemental Material, Section III.B). 

The typical ground state (GS) energy $E$ (\ref{erbm}) of R-RBM can be computed with the replica method within the replica-symmetric Ansatz \cite{ags}, as the optimum of
\begin{eqnarray}\label{egs}
E_{GS} & =& \frac L2 m^2 + \frac \alpha 2  (q\, B + r\,C) - \frac 1N \sum_i\sqrt{\alpha\, p_i\, r}  \\
&\times &\left\langle H^{(1)} \left( -\bigg[ g +\frac \alpha 2 B\, p_i+ m \,W \bigg]/\sqrt{\alpha \,p_i\,r}\right)
\right\rangle_{W} \nonumber \\
& + & \alpha \int Dz\; \min _h\bigg( {\cal U}^{ReLU}(h) - \frac {C}2 \, h^2 -z \,\sqrt{p q}\,h \bigg)\nonumber 
\end{eqnarray}
over the order parameters $m,L,r,q,B,C$ (averaged over the quenched weights): $m$ and $L$ are, respectively, the magnetization and the number of feature-encoding hidden units, $r$ is the mean squared activity of the other hidden units, $q=\sum_i p_i \langle v_i\rangle_{GS}/(Np)$ is the weighted activity of the visible layer in the GS, and $B,C$ are response functions, {\em i.e.} derivatives of the mean  activity of, respectively, hidden and visible units with respect to their inputs \cite{long}.  In (\ref{egs}) $Dz=\frac {dz}{\sqrt{2\pi}} e^{-z^2/2}$ denotes the Gaussian measure, $H^{(k)}(x)=\int _x Dz (z-x)^k$, and $\langle\cdot\rangle_W$ is the average over the sum $W$ of $L$ i.i.d. weights $w_{i\mu}$ drawn as above.

\begin{figure}[h]
\begin{center}
\includegraphics[width=1\linewidth]{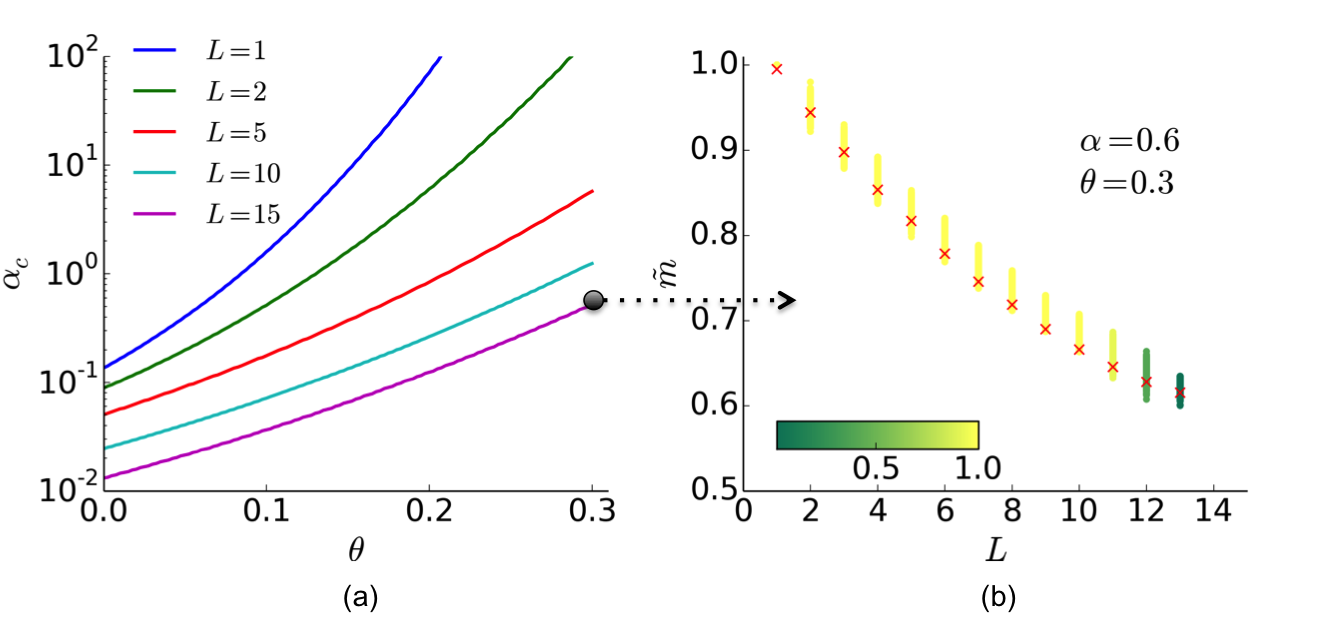}
\end{center}
\caption{Compositional regime in R-RBM. {\bf (a)} Critical lines in the $\theta,\alpha$ plane below which $L$ hidden units may be strongly activated, calculated from the optimization of $E_{GS}$ (\ref{egs}).  Parameters are $p_i=p=0.1$, $g= -0.02$. {\bf (b)} Comparison of theoretical (red crosses) and numerical simulations ($N= 10^4$ visible units, colored points) for the rescaled magnetizations $\tilde m=m/(p/2))$ as a function of the number $L$ of strongly activated hidden units in R-RBM. 7,500 zero temperature MCMC, each initialized with a visible configuration strongly overlapping with $L=1,2,...$ `features', were launched; color code indicates the probability that the same $L$ hidden units are magnetized after convergence (see bottom scale),  and the corresponding average magnetization $\tilde{m}$. }
\label{fig:compo}
\end{figure}

We first fix $L$, and optimize $E_{GS}$ over all the other order parameters. At large $\alpha$ the only solution has $m=0$, and corresponds to the Spin-Glass phase. For intermediate values of $\alpha$, other solutions, with $m>0$, exist.  For the sake of simplicity we consider first the homogenous sparsity case, with $p_i=p$. We show in Fig.~\ref{fig:compo}(a), for fixed $p=0.1$ and various values of $L$, the maximal value of $\alpha$ below which a phase with $L$ magnetized hidden units exists. Importantly this critical value can be made arbitrarily large by increasing the ReLU threshold $\theta$. This phenomenon is a consequence of the nonlinearity of ReLU, and can be understood as follows. The squared activity $r$ of non-magnetized hidden units obeys the saddle-point equation $r= p q/(1-C)^2 \times H^{(2)}(\theta/\sqrt{p q})$. The first factor is reminiscent of the expression $r=1/(1-C)^2$ arising for the Hopfield model (for which $p=q=1$ at zero temperature) \cite{ags}, while the second factor comes from the nonlinearity of ReLU. $H^{(2)}$ being a rapidly decaying function of its argument large thresholds $\theta$ lead to small $r$ values. As the level of crosstalk due to nonmagnetized hidden units diminishes  larger ratios $\alpha$ can be supported by R-RBM without entering the glassy phase. Numerical simulations of R-RBMs at  large $\alpha$ confirm the existence and (meta)stability of phases with $L$ nonzero magnetizations (Fig.~3(b)). Moreover, the values of the average normalized magnetizations $\tilde m = m/(p/2) \in [0;1]$ are in excellent agreement with those found by optimizing $E_{GS}$.

The nature of the large--$L$ phases and the selection of the value of $L$ are best understood in the limit case of  highly sparse connections, $p\ll 1$. The R-RBM model exhibits an interesting limit behaviour, which we call hereafter {\em compositional} phase. In this regime the number of strongly magnetized hidden units is unbounded, and diverges as $L=\ell/p$, with $\ell>0$ and finite. The normalized GS energy $e_\ell = E_{GS} (L=\ell/p)/p$ is a nonmonotonous function of the index $\ell$, see Fig.~4(a). Minimization of $e_\ell$ leads to the selection of a well defined index $\ell^*$. The magnetizations of the $\ell^*/p$ strongly activated units, $m=\frac p2\, \tilde m$, vanish linearly with $p$ \footnote{Solutions  with nonhomogeneous magnetizations $m_\mu$, varying from one strongly activated hidden unit to another, give additional contributions to $E_{GS}$ of the order of $p^2$ with respect to the homogeneous solution $m_\mu=m$, and do not affect the value of $e_\ell$ \cite{long}.}. Nonmagnetized hidden units have activities of the order of $\sqrt r \sim \sqrt p$, and can be shutdown by choosing thresholds $\theta \sim \sqrt p$; hence crosstalk between those units can be suppressed, allowing for large relative size $\alpha$ of the hidden layer. The input received by a visible unit from the large number of magnetized units is, after transmission through the dilute weights, of the order of $L\, m \, p =\frac 12\,\ell^*\,\tilde m\, p$; it can be modulated by a (positive or negative) field $g \sim p$ to produce any finite activity $q$ in the visible layer, as soon as the effective temperature gets below  $\sim p$. 

The compositional phase competes with the ferromagnetic phase, in which $\tilde m>0$ but $e_\ell$ is a monotonously growing function of $\ell$ (hence, $\ell^*=0$), and the spin glass phase, in which $\tilde m=0$ and $e_\ell$ does not depend on $\ell$, see Fig.~4(a). The  phase diagram in the parameter space ($\alpha,\tilde g = \frac{g}{p},\tilde \theta = \frac{\theta}{\sqrt{p}})$ will be detailed in \cite{long}. Briefly speaking, given $\alpha$, $\tilde \theta$ should be large enough (as in Fig.~3) and $|\tilde g|$ should be neither too large to penalize the ferromagnetic phase, nor too small to avoid the spin glass  regime. 

\begin{figure}[h]
\begin{center}
\includegraphics[width=1\linewidth]{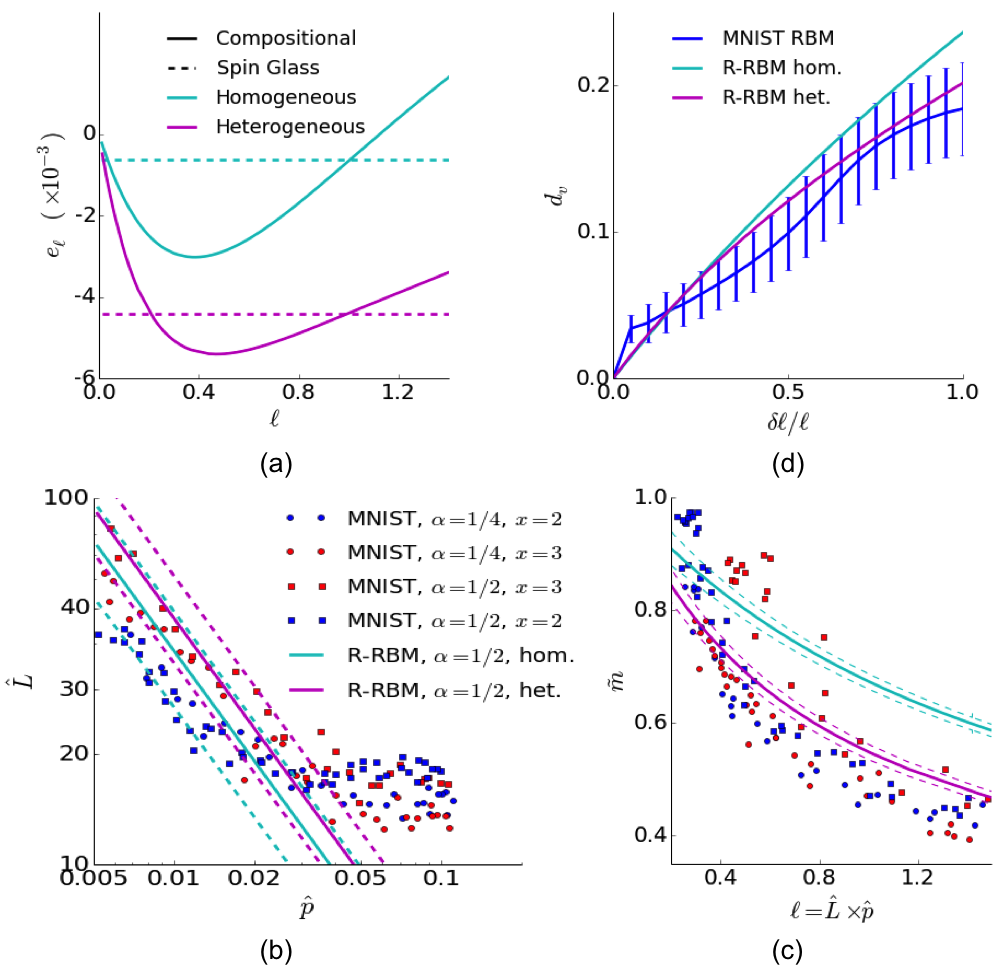}
\end{center}
\caption{{\bf (a)}. Behaviour of the GS energy $e_\ell$ vs. $\ell = L \times p$ in the $p\to 0$ limit. Parameters: $\tilde \theta =1.5$, $\alpha = 0.5$. 
{\bf (b)}-{\bf (c)}-{\bf (d)}. Quantitative predictions in the compositional regime of R-RBM compared to RBMs inferred on MNIST. Each point represent a ReLU RBM trained with various regularizations, yielding different weight sparsities $p$. Solid lines depict predictions found by optimizing $e_\ell$ (\ref{egs}), and dashed line expected fluctuations at finite size ($N$) and temperature.
{\bf (b)}  Average number $L$ of active hidden units  vs.  $p$. 
{\bf (c)}  Average magnetization $\tilde m$ vs. $\ell = L \times p$.
{\bf (d)} Distance (per pixel) between the pairs of visible configurations after convergence of zero-temperature MCMC vs. relative distances in the hidden-unit activation patterns. MCMC are initialized  with all pairs of digits in MNIST; final visible configurations differ from MNIST digits by about 7 pixels,  both on the training and test sets, see Supplemental Material, Fig.~1.
In all four panels  predictions for the homogeneous ($\tilde g =  - 0.21$) and  heterogeneous  ($\tilde g = - 0.1725$, see Supplemental Material, Section III.E) cases are shown in, respectively, cyan and magenta.}
\label{fig:scaling}
\end{figure}


Characteristic properties of our compositional phase are confronted to ReLU RBMs trained on MNIST in Fig.~4 (b,c). Compared to Fig.~2 we add a regularization penalty $\propto\sum_{\mu} ( \sum_i |w_{i\mu} | )^x$ to control the final degree of sparsity; the case $x=1$ gives standard $L_1$ regularization, while, for $x>1$, the effective penalty strength $\propto (\sum_i |w_{i\mu}|)^{x-1}$  increases with the weights, hence promoting homogeneity among hidden units. After training we generate Monte Carlo samples of each RBM at equilibrium, and monitor the average number of active hidden units, $\hat{L}$, and the normalized magnetization, $\tilde{m}$. Figure~4(b) shows $\hat{L}$ vs. $\hat{p}$, in good agreement with the R-RBM theoretical scaling $L \sim \frac{\ell^\star}{p}$. Figure~4(c) shows that $\tilde{m}$ is a decreasing function of  $\ell = \hat{L} \times \hat{p}$, as qualitatively predicted by theory, but quantitatively differs from the prediction of R-RBM with homogeneous $p$. This disagreement can be partly explained by the heterogeneities in the sparsities $p_i$ in RBMs trained on MNIST, {\em e.g.} units on the borders are connected to only few hidden units, whereas units at the center of the grid are connected to many. We introduce a heterogeneous R-RBM model, where the distribution of the $p_i $'s is fitted  from MNIST-trained RBMs (Supplemental Material, Section III.E). Its GS energy can be calculated from (\ref{egs}), see Fig.~4(a) \cite{long}. Results are shown in Fig.~4(b,c) to be in good agreement with RBM trained on MNIST. 

RBMs, unlike the Hopfield or mixture model, may produce gradually different visible configurations through progressive changes in the hidden-layer activation pattern. R-RBMs enjoy the same property. We compute, through a real-replica approach \cite{long}, the average Hamming distance $d$ (per pixel) between the visible configurations ${\bf v}^{(1)},{\bf v}^{(2)}$ minimizing the energy $E$ (\ref{erbm}) for two hidden configurations ${\bf h}^{(1)},{\bf h}^{(2)}$  sharing $(\ell-\delta\ell)/p$ hidden units among the $\ell/p$ strongly activated ones. Figure~4(d) shows that $d$ monotonously increases from $d=0$ for $\delta\ell=0$ up to $d=2q(1-q)$ (complete decorrelation of visible units) for $\delta \ell=\ell$, in very good quantitative agreement with results for RBM trained on MNIST. 

The gradual change property has deep dynamical consequences. MCMC of MNIST-trained RBM (videos available in Supplemental Material) show that gradual changes may occasionally lead to another digit type, by passing through well-drawn, yet ambiguous digits. The progressive replacement of feature-encoding hidden units (small $\delta \ell$ steps)  along the transition path does not increase much the energy, and the transition process is fast compared to activated hopping between deep minima taking place in the Hopfield model. 

Our study is related to several previous works. RBMs with linear activation function $\Phi$ coincide with the Hopfield model. In this framework magnetized hidden units identify retrieved patterns, and $\alpha$ corresponds to the capacity of the autoassociative memory. Tsodyks and Feigel'man showed how the critical capacity (for single pattern retrieval) could be dramatically increased with sparse weights ($p\ll 1$) and appropriate tuning of the fields $g_i$ \cite{tsodyks}; however this effect could be achieved only with vanishingly low activities $q$. Agliari and collaborators showed in a series of papers \cite{barra,barra2} that multiple sparse patterns could be simultaneously retrieved in the case of linear $\Phi$ and vanishing capacity $\alpha=0$ (finite $M$). Finite capacity $\alpha\sim c^{-2}$ could be achieved at zero temperature in the limit of extreme sparsity,  $p=c/N$, only \cite{sollich}; for MNIST $p\simeq 0.1$ and $N=784$ would give $\alpha\sim 2.10^{-4}$. Our work shows that large values of $\alpha$ can be reached even with moderate sparsity $p$ (as in realistic situations, see Fig.~2) provided that nonlinear $\Phi$ (ReLU) and appropriate threshold values $\theta$ are considered. The presence of the fields $g_i$ acting on the visible units (absent in the $v_i=\pm1$ model of \cite{barra,barra2,sollich}), is also crucial for the existence of our compositional phase as explained above.

It would be interesting to extend our work to more than one layers of hidden units, or to other types of nonlinear $\Phi$. While numerical studies of RBMs with Bernoulli hidden units show no qualitative change compared to ReLU, choosing $\Phi(h)$ growing asymptotically faster than $h$ could affect the nature of the extracted features \cite{krotov}. An important challenge would be to understand the training dynamics, {\em i.e.} how hidden units gradually extract features from data prototypes.


\vskip .2cm\noindent
{\bf Acknowledgements.} We are grateful to C. Fisher and G. Semerjian for useful discussions. This work was partly funded by the CNRS-Inphyniti Inferneuro and the HFSP  RGP0057/2016 projects, and benefited from the support of NVIDIA Corporation with the donation of a Tesla K40 GPU card.

\end{document}


\begin{center}
  \LARGE Supplemental Material
\end{center}
\title{Emergence of Compositional Representations in Restricted Boltzmann Machines}
\author{J. Tubiana, R. Monasson}
\maketitle

\section{Training RBMs on MNIST}
\subsection{Dataset preparation and initial conditions}
\begin{itemize}
\item In MNIST, each pixel has a value between 0 and 255. We binarize it by thresholding $\geq 128$. The $28 \times 28$ binary images are flattened to a $N = 784$ vector with binary values.
\item The dataset is split in a training (60,000 instances) and a test (10,000 instances) sets
\item The weights $w_{i\mu}$ are randomly initialized at $\pm W$, where $W = \sqrt{\frac{0.1}{N}}$; this choice corresponds to initial temperature and weight sparsity : $T(0) = 10$ and $p(0) = 1$ (see section III).
\item The initial field values are $g_i^0 = \log \left[ \frac{\langle v_i\rangle^{MNIST}}{1-\langle v_i\rangle^{MNIST}} \right]$, where $\langle v_i\rangle^{MNIST}$ denotes the average of pixel $i$ over the training data
\item For ReLU, the thresholds $\theta_\mu$ are all initially set to $0$
\end{itemize}

\subsection{Learning algorithms}
A RBM is associated to a probability distribution $P[{\bf v},{\bf h}) = \frac{e^{-E[{\bf v},{\bf h}]}}{Z}$, where the energy $E$ is defined in the main text. The marginal distribution, $P[{\bf v}] = \int \prod dh_\mu P[{\bf v},{\bf h})$ is fitted to the data by likelihood maximization. Given data instances ${\bf x}^{r}, r \in \{1,D\}$, the log-likelihood is :

\begin{equation}
\log \mathcal{L}_{{\bf W}, {\bf g}, \bm{\theta}} = \frac{1}{D} \sum_{r=1}^D \log \left[ P[{\bf x}^r | {\bf W}, {\bf g}, \bm{\theta}|  \right]
\end{equation}

Where ${\bf W}$ is the matrix of weights, ${\bf g}$ is the vector of visible layer fields and $\bm{\theta}$  is the vector of hidden units thresholds. Likelihood maximization is implemented by stochastic gradient descent, with the difficulty that extensive Monte Carlo simulations are required to compute the gradient \cite{igel,training1}. For the RBM of Fig. 2 in the main text, we used Persistent Contrastive Divergence \cite{training2} with 
\begin{itemize}
\item 20 mini-batch size
\item 100 persistent chains
\item 1 Gibbs step between each update
\item 200 epochs (600 000 updates in total)
\item Initial learning rate of $\lambda_i = 5 \; 10^{-3}$, decaying geometrically (decay starts after 60 epochs) to $\lambda_f = 5 \;10^{-4}$
\end{itemize}

PCD is known to be inaccurate toward the end of learning, because the parameters evolve too fast with respect to the the mixing rate of the Markov chains. The regularized RBM of main text, Fig. 4 (b,c), were trained with a more efficient algorithm, variant of Adaptive Parallel Tempering \cite{training3,long} with 
\begin{itemize}
\item 100 mini-batch size
\item 100 persistent chains
\item 10 replicas
\item 1 Gibbs step between each update
\item 150 epochs (90 000 updates in total)
\item Initial learning rate of $\lambda_i = 10^{-2}$, decaying geometrically (decay starts after 90 epochs) to $\lambda_f = 10^{-4}$
\end{itemize}

\subsection{Monitoring the learning}
We monitor the evolution of the likelihood and of the pseudo-likelihood of the train and test data sets throughout learning, see Fig. 1(a). The choice of parameters made learning slow, but ensured that the likelihood increased steadily throughout training. The likelihood requires approximate computation of the model partition function; Annealed Importance Sampling \cite{AIS} was used. Parameters :  $n_\beta = 10000$ inverse temperatures with an adaptive spacing \cite{long}, $n_{runs} =1$. 
Additionnaly, we can look at the probability landscape $P_W(v)$ throughout learning. For each of the 70k MNIST samples, a gradient ascent on $P_W(v)$ is performed until convergence to a local maximum; the number of distinct local maxima of $P_W(v)$ and the distance to the original sample are measured. As training goes, more local maxima appear, and they get closer to the training samples; local maxima also appear close to the test set, which shows that RBM generalize well.

\begin{figure}
\begin{subfigure}{.5\textwidth}
  \centering
  \includegraphics[scale=0.35]{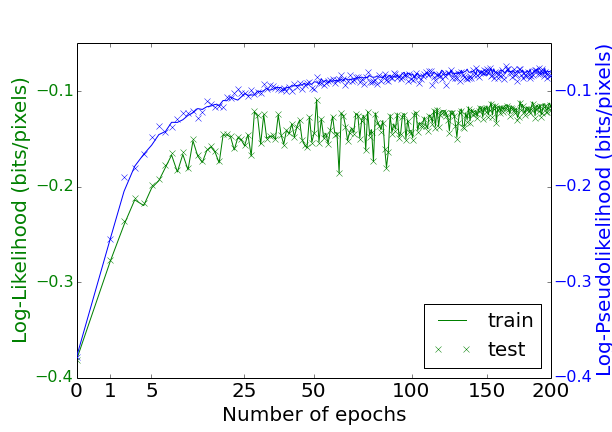}
\end{subfigure}%
\begin{subfigure}{.5\textwidth}
  \centering
  \includegraphics[scale=0.35]{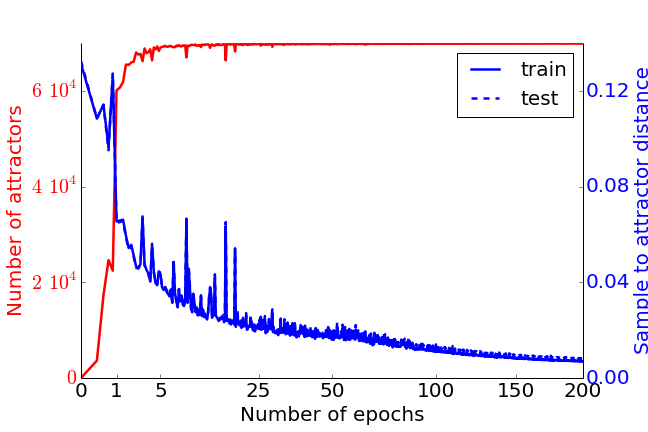}
\end{subfigure}
\caption{ {\bf (a)} Evolution throughout training of the data loglikelihood (left scale) and pseudo-loglikelihood (right scale) computed over the training and test sets. {\bf (b)} Evolution of the number of distinct local maxima of $P_W(v)$ (left scale) and the distance to the original sample (right scale, for train and test set) are displayed.}
\end{figure}

\subsection{Controling weight sparsity with regularization}
To control the weight sparsity $p$, a regularization penalty is added to the log likelihood $\log \mathcal{L}_{{\bf W}, {\bf g}, \bm{\theta}}$:

\begin{equation}
\begin{split}
\text{Cost} &= -\log \mathcal{L}_{{\bf W}, {\bf g}, \bm{\theta}}+ L^{(x)} \\
 L^{(x)} &= \frac{\lambda_x}{x} \sum_\mu \left[\sum_i |w_{i \mu} | \right]^x \\
 -\frac{\partial \text{Cost}}{\partial w_{i \mu}} &=  \frac{\partial}{\partial w_{i \mu}} \log \mathcal{L}_{{\bf W}, {\bf g}, \bm{\theta}} - \lambda_x \left[ \sum_j |w_{j \mu}| \right]^{x-1} \text{sign}(w_{i \mu})
\end{split}
\end{equation}

The case $x = 1$ is the usual $L_1$ penalty and performing gradient descent with $\lambda_1>0$ is known to reduce the number of non-zero weights $w_{i\mu}$. However, experiments show that the outcome is inhomogeneous with respect to the hidden units: some hidden units are weakly affected by the penalty, whereas some end up completely disconnected from the visible layer, making them useless, see Fig. 2. To maintain homogeneity among the hidden units, we pick $x = 2$ or $x=3$. As can be seen from the expression of the gradient, it is equivalent to a usual $L_1$ penalty, but with a decay rate adaptive to each hidden unit: hidden units strongly (resp. weakly) coupled to the visible layer (large $\sum_i |w_{i \mu}|$) are strongly (resp. weakly) regularized, thus increasing the homogeneity among hidden units.

\begin{figure}
\includegraphics[scale=0.7]{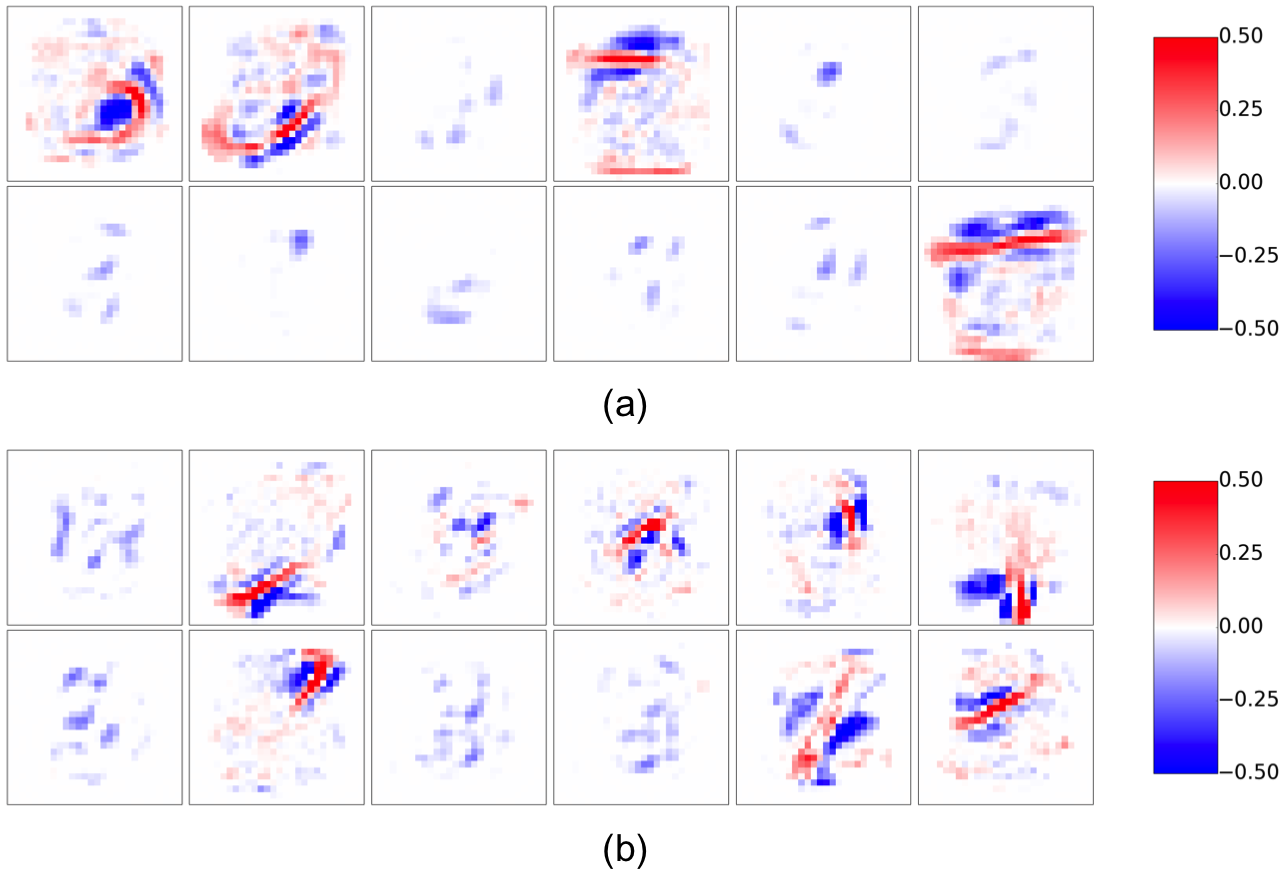}
\caption{Subset of 12 weight features produced by training on MNIST, regularized with $L^1, \lambda_1 = 10^{-3}$ (top panel), and $L^2, \lambda_2 = 3 . 10^{-5}$ (bottom panel). Both have overall sparsity $p \sim 0.036$, but the latter is more homogeneously distributed across hidden units.}
\end{figure}

\section{Sampling from RBMs}

RBM can be sampled by Markov Chains Monte Carlo. Due to the conditional independence property, Gibbs sampling can be performed by alternative sampling of $\bf h$ from $P[{\bf h}|{\bf v}]$, then $\bf v$ from $P[{\bf v}|{\bf h}]$ \cite{training1,igel}.

\subsection{MCMC Videos}
The two videos in Supplementary  Material present visualize MCMC runs from RBM trained on MNIST with Bernoulli, Gaussian, ReLU hidden units. Each square depicts a Markov chain started from a random initial condition. 20 Gibbs steps are performed between each image, and each chain is 500 images long. See Fig. 3 for a snapshot.

\begin{figure}
\includegraphics[scale=0.7]{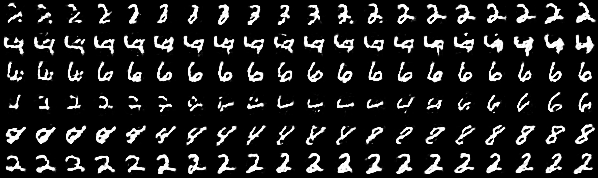}
\caption{Six independent Monte Carlo Markov Chains realization for a RBM trained on MNIST, extracted from the attached videos, see text.}
\end{figure}

\subsection{Estimating thermal averages with MCMC}
Sampling at thermal equilibrium is required to estimate the values of order parameters ($L$, $S$, $q$, $\tilde{m}$,...). Since RBM trained on MNIST effectively operate at low temperature (entropy of 0.1 bits/pixel) the MCMC mixing rate is poor, and long simulations would be required for each of the $\sim 100$ RBMs trained. To overcome this issue we use an Adaptive Parallel Tempering (also known as Replica Exchange) sampling algorithm, with 10 replicas \cite{training3,long}. Observables are averaged over 100 independent Markov Chains, each being first thermalized for 500 Gibbs updates, then run for another 100 Gibbs updates (10K samples in total).

\begin{figure}
\includegraphics[scale=0.7]{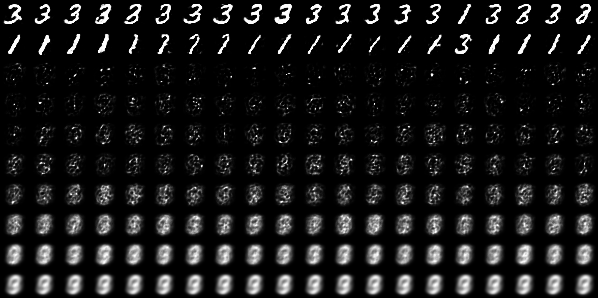}
\caption{Ten Monte Carlo Markov Chains realizations at different inverse temperatures, coupled by replica exchange. The plots shows the conditional expectations of visibile units, $\mathbb{E} \left[ {\bf v} | {\bf h} \right]$, for thermalized hidden-unit activities, ${\bf h}$.}
\end{figure}

\subsection{Estimating order parameters of R-RBM with zero temperature MCMC}
R-RBM are studied analytically in the zero-temperature limit; this limit can be simulated as well. The energy $E[{\bf v},{\bf h}]$ of a configuration ${\bf v},{\bf h}$ is given by Eqn. (1) in main text, and defines  the Gibbs distribution $P^\beta[{\bf v},{\bf h}] = \exp(- \beta E[{\bf v},{\bf h}]/Z(\beta)$, 
where $\beta = \frac{1}{T}$ is the inverse temperature. As $\beta$ increases, $P^\beta[{\bf v},{\bf h}]$ is more and more peaked around the minimum of $E$. In the limit $\beta \rightarrow \infty$, a dynamical Gibbs step becomes deterministic :

\begin{equation}
\begin{split}
&h_\mu \leftarrow \arg\min_h \left[ U_\mu(h) - h \sum_i w_{\mu i} v_i \right] \equiv \Phi_\mu \left[ \sum_i w_{i\mu } v_i  \right] \\
&v_i \leftarrow  \arg\min_v \left[ - g_i v - v \sum_\mu w_{\mu i} h_\mu   \right] \equiv  \Theta \left[ g_i +  \sum_\mu w_{i\mu } h_\mu \right]\ ,
\end{split}
\end{equation}
where $\Theta$ is the Heaviside function, and $\Phi_\mu$ is the response function (Fig.~1(b) in main text).Starting from a configuration, such zero-temperature Markov Chain runs until convergence to a local minimum of $E$.

In practice, to make finite-size corrections to our mean-field theory small, we considered RBMs with up to $N \sim 10^4$ visible units. Such large R-RBM were simulated using a nVidia Tesla K40 GPU, programmed with Theano \cite{theano}.

\subsection{Finding local maxima of $P[{\bf v}]$}
Given an RBM with energy defined as above, the marginal $P[ {\bf v}]$ is characterized by a Gibbs distribution and a free energy :
\begin{equation}
\begin{split}
P[{\bf v}] =\int \prod_\mu dh_\mu \frac{1}{Z} e^{-E[{\bf v}, {\bf h}]} =  \frac{1}{Z} e^{-F[{\bf v}]} \\
F[{\bf v}] = - \sum_i g_i v_i + \sum_\mu U_\mu^{eff}\left( \sum _i w_{i\mu} v_i \right)  \\
U_\mu^{eff}(x) = -  \log \left[ \int dh \; e^{-U_\mu(h) + x h} \right]
\end{split}
\end{equation}

In order to find the local maxima of $P[{\bf v}]$ (\textit{i.e.} the local minima of $F[{\bf v}]$ ) , we modify it by introducing an inverse temperature $\beta$: 
\begin{equation}
\begin{split}
P^\beta[{\bf v}] =\frac{1}{Z(\beta)} e^{-\beta F[{\bf v}]} \\
Z(\beta) = \sum_{\bf v} e^{-\beta F[{\bf v}]}
\end{split}
\end{equation}

Sampling from this distribution at $\beta \neq 1$ is not trivial, as $P^\beta [{\bf v}] $ is not the marginal distribution of $P^\beta[{\bf v},{\bf h}]$ when $\beta\ne 1$.
While sampling from $P^1[{\bf v}]$ is easy, as one can simply sample from the joint distribution $P^1[{\bf v}, {\bf h}]$ using Gibbs steps, this is not true for $\beta \ne 1$; in particular the local maxima of $P^\beta[{\bf v},{\bf h}]$ are not  equivalent to those of $P^\beta[{\bf v}]$. We notice however that when $\beta \geq 1$ is an integer, $P^\beta [{\bf v}]$ can be interpretated as the $\beta = 1$ distribution of another RBM $P'^1[{\bf v}]$ with $\beta M$ hidden units (each hidden unit is replicated $\beta$ times) and visible fields $g' = \beta g$.

Sampling from such RBM can be done as following :
\begin{itemize}
\item Compute the hidden layer inputs $\sum _iw_{i\mu} v_i$
\item Sample independently $\beta$ replicas $h_\mu^r$  of $h_\mu$ from $P^1[ h_\mu | {\bf v}]$
\item Compute the visible layer inputs $I_i =\sum_{r=1}^\beta \sum _\mu w_{i\mu} h^r_\mu $
\item Sample $v_i$ from the Bernoulli distribution $Bern \left[ \beta ( g_i + \frac{1}{\beta}I_i ) \right] $

\end{itemize}

When $\beta \rightarrow \infty$, $\frac{1}{\beta}\sum_{r=1}^\beta h_r$ coincides with the conditional average of ${\bf h}$ given ${\bf v}$,  $\mathbb{E}\left[ {\bf h} | {\bf v} \right]$. Therefore, the zero temperature sampling Gibbs step for the free energy is equivalent to :
\begin{equation}
\begin{split}
&h_\mu \leftarrow \mathbb{E}\left[ h_\mu | v \right] \\
&v_i \leftarrow \Theta \left[ g_i + \sum_\mu w_{i\mu} h_\mu \right]
\end{split}
\end{equation}

\section{Numerical proxies for control and order parameters}
Several control and order parameters are well defined for R-RBM in the thermodynamical limit, but not for typical RBM trained on data. For R-RBM instances, the average weight sparsity $p$ is well defined because the weights take only three distinct values $\{-\frac{1}{\sqrt{N}},0,\frac{1}{\sqrt{N}} \}$, but for RBM trained on data, the weights $w_{i\mu}$ are never exactly equal to zero. Similarly, the number of strongly activated hidden units $L$ is well-defined for R-RBM in the thermodynamic limit $N \rightarrow \infty$ because their activity scales as $\sqrt{N}$; but in general, all hidden units have finite activation. Proxies are therefore required to compare theoretical and numerical results. We consider 'consistent' proxies, giving back (in the large size limit), the original parameters for RBMs drawn from the R-RBM ensemble.

\subsection{Participation Ratios $PR$}
Participation ratios are used to estimate numbers of nonzero components in a vector, while avoiding the use of arbitrary thresholds. The Participation Ratio $(PR_a)$ of a vector ${\bf x}=\{x_i \}$ is $$PR_a(\bf x) =  \frac{(\sum_{i} |x_i|^a)^2}{\sum_{i} |x_i|^{2a} }$$ If $\bf x$ has $K$ nonzero and equal (in modulus) components PR is equal to $K$ for any $a$. In practice we use the values $a=2$ and 3: the higher $a$ is, the more small components are discounted against strong components in $\bf x$.

\subsection{Number $L$ of active hidden units}

In both numerical simulations of R-RBM and on RBM trained on MNIST, we estimate $L$, for a given hidden-unit configuration $\bf{h}$, through 
$$\hat{L}  = PR_3(\bf{h})$$

To understand the choice $a=3$, consider a typical activation configuration $\bf{h}$ for a R-RBM : 
\begin{equation}
h_\mu = \left\{ \begin{array}{c c c}
m \sqrt{N} & \text{if } &1\le \mu\le L \ , \\
\sqrt{r}\; x_\mu & \text{if }& L+1\le  \mu \le M\ , 
\end{array} \right.
\end{equation}
where the magnetization $m$ and mean square activity $r$ are $\mathcal{O}(1)$, and $x_\mu$ are random variables with zero mean, and even moments of the order of unity. The first $L$ hidden units are strongly activated ($\mathcal{O}(\sqrt{N})$ activity), whereas the remaining $N-L$ others are not (activations of the order of 1). Here, we assume $L$ to be finite as $N\to\infty$. One computes : 
\begin{equation}
\begin{split}
PR_2(h) \sim \frac{(L m^2 N  + (N-L) r)^2}{L m^4 N^2 + (N-L) r^2} = L \times \frac{(1+\frac{N-L}{N} \frac{r}{L m^2})^2}{1 + \frac{N-L}{N^2} \frac{r^2}{Lm^4}}\underset{N \rightarrow \infty}{\longrightarrow} L (1 + \frac{r}{L m^2})^2 \ ,\\
PR_3(h) \sim \frac{(L m^3 N^{3/2}  + (N-L) r^{3/2})^2}{L m^6 N^{3} + (N-L) r^3} = L \times \frac{(1+\frac{N-L}{N^{3/2}} \frac{r^{3/2}}{L m^3})^2}{1 + \frac{N-L}{N^3} \frac{r^3}{Lm^6}} \underset{N \rightarrow \infty}{\longrightarrow} L\ .
\end{split}
\end{equation}
Hence choosing coefficient $a=3$ ensures that the participation ratio (a) does not take into account the weak activations in the thermodynamical limit, and (b) converges to the true value $L$ if all magnetizations are equal.

\subsection{Normalized Magnetizations $\tilde{m}$}
Given a RBM and a visible layer configuration, we define the normalized magnetization of hidden unit $\mu$ as the normalized overlap between the configuration and the weights attached to the unit: $$\tilde{m}_\mu = \frac{\sum_i (2 v_i -1) w_{i\mu}}{\sum_i | w_{i\mu} | } \in [-1,1]$$ This definition is consistent with the Hopfield model. For R-RBM, we also have, in the thermodynamical limit, $\hat{m}_\mu = \frac{2I_\mu}{p \sqrt{N}} $, where $I_\mu$ is the input received by the hidden unit from the visible layer; $m_\mu$ is $\mathcal{O}(1)$ for strongly activated hidden units, and $\mathcal{O}(\frac{1}{\sqrt{N}} )$ for the others.

For a given configuration $\bf{v}$, with $\hat{L}$ activated hidden units, the normalized magnetization of the activated hidden units $\tilde{m} = \frac{m}{p/2 }$  can be estimated as the average of the $\hat{L}$ highest magnetizations $\hat{m}_\mu$.

\subsection{Weight sparsity $p$}
A natural way to estimate the fraction of non-zero weights $w_{i\mu}$ would be to count the number of weights with absolute value above some threshold $t$. However, there is no simple satisfactory choice for $t$. Indeed, the fraction of non-zero weights should not depend on the scale of the weights, {\em i.e.} it should be invariant under the global rescaling transformation $\{w_{i\mu}\} \rightarrow \{\lambda\, w_{i\mu}\}$. As the scale of weights vary from RBMs to RBMs and, for each RBM, across training it appears difficult to select an appropriate value for $t$. A possibility would be to use a threshold adapted to each RBM, {\em e.g.} $t \propto \kappa \sqrt{\frac{W_2}{M}}$, where $\kappa$ would be some small number. Our experiments show that it is not accurate enough, due to the scale disparities across the hidden--unit weight vectors ${\bf w}_\mu$. Rather than adapting thresholds to each hidden unit of each RBM, we use Participation Ratios, which naturally enjoy the scale invariance property. We estimate the fraction of nonzero weights through
$$ \hat{p} = \frac{1}{MN} \sum_\mu PR_2({\bf w}_\mu )$$
For R-RBM with $w_{i \mu} \in [-W_0, 0, W_0]$ with corresponding probabilities $[\frac p2 , 1-p,\frac p2]$, the estimator is consistent: $\hat{p} = p$. 


\subsection{Weights heterogeneities}
\begin{figure}
\includegraphics[scale=0.7]{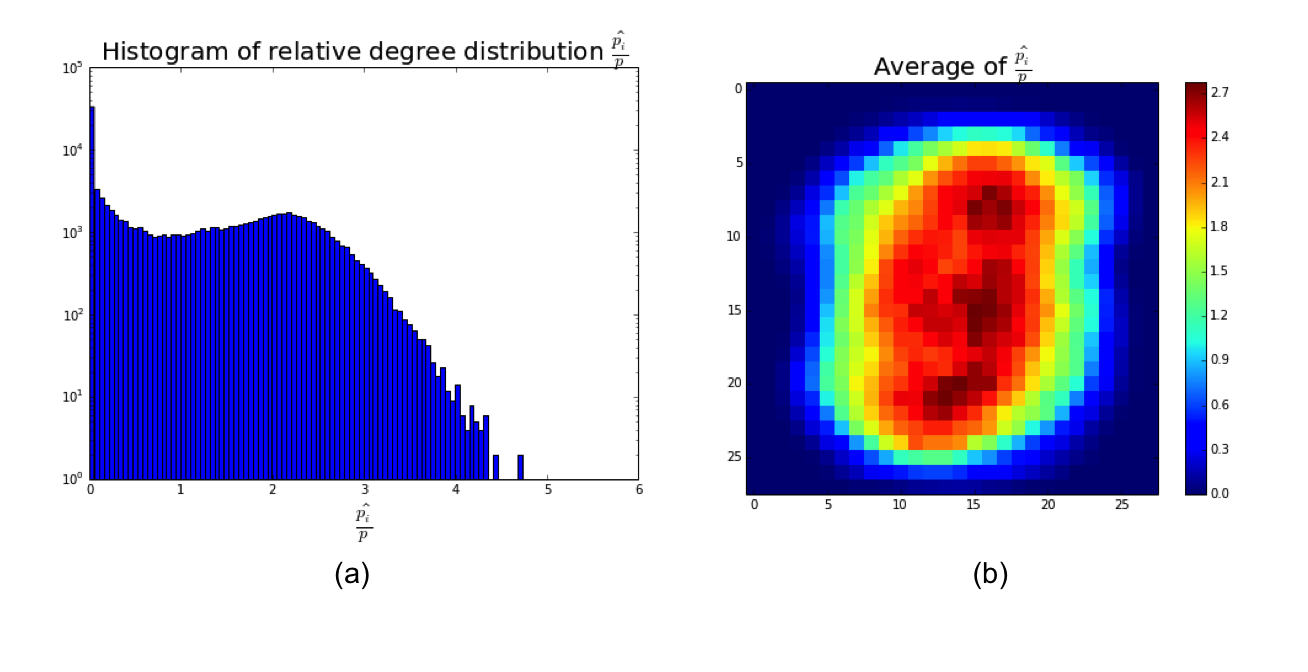}
\caption{(a) Histogram of $\tilde p_i = \frac{p_i}{p}$ values, across all visible units and RBMs inferred on MNIST. (b) Average across all RBM of $\tilde p_i =\frac{p_i}{p}$, for each visible unit}
\label{distpt}
\end{figure}

As seen from the features of Fig.~2 in the main text, not all visible units are equally connected to the hidden layer. To better capture this effect, one can study R-RBM with any arbitrary distribution of $p_i$. Analogously to the homogeneous case  a high sparsity limit is obtained when the average sparsity, $p=\frac 1N \sum _i p_i$, vanishes. We define the distribution of the ratios $\tilde p_i=\frac{p_i}{p}$ in the $p\to 0$ limit. In practice the ratios are estimated through 
\begin{equation}
\tilde p_i = \frac{\sum_i w_{i\mu}^2}{\frac{1}{M} \sum_{i,\mu} w_{i\mu}^2}\ .
\end{equation}
For a heterogeneous R-RBM, we have consistently $\tilde p_i = \frac{\hat p_i}{p} = \frac{p_i}{p}$. Looking at the histogram of values of $\tilde p_i$ across all RBM inferred on MNIST, we find a non-negligible spread around one, see Fig.~\ref{distpt}. We also display for each visible unit $i$ the average of $\tilde p_i$ accross all RBM inferred; we can see that the visible units at the border are indeed the least connected (smaller $\tilde p_i$), whereas the ones at the center are strongly connected (larger $\tilde p_i$).

\subsection{Effective Temperature $T$}

\begin{figure}
\includegraphics[scale=0.6]{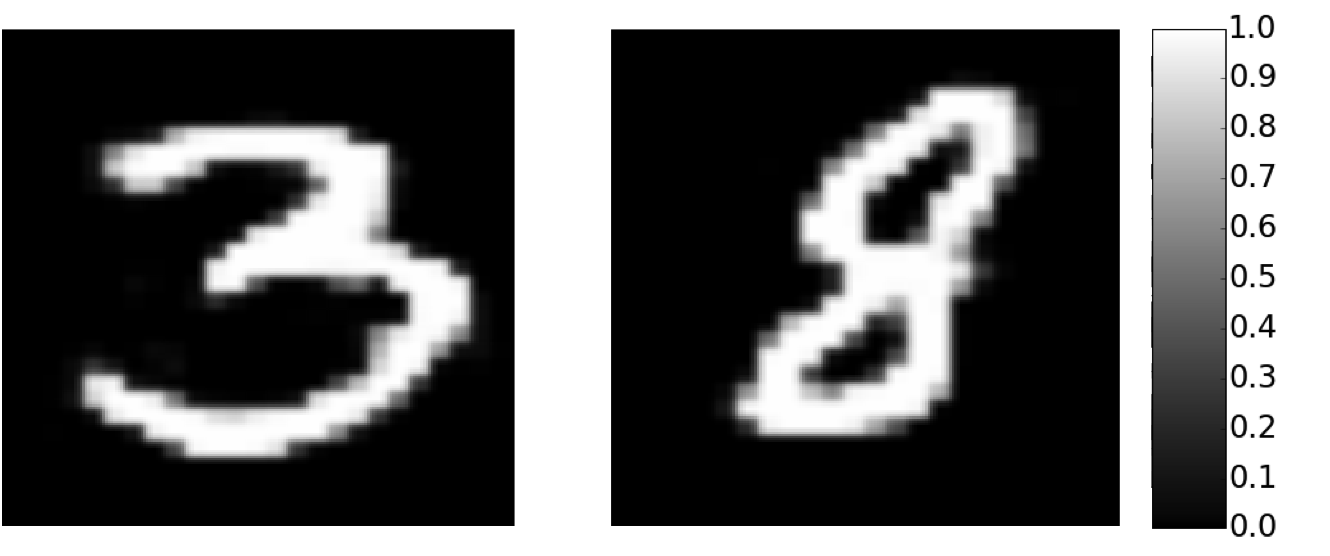}
\caption{Conditional means $\mathbb{E}\left[{\bf v} | {\bf h} \right]$ for two hidden units configurations sampled at equilibrium. Most pixels $v_i$ are frozen, with $\mathbb{E}\left[v_i | {\bf h} \right] \in \{0,1\}$}
\end{figure}

Although RBM distributions are always defined at temperature $T=1$, the effective temperature is not $1$. This is very much like in the Ising model : the behavior of the system depends on an effective temperature $\hat{T} = \frac{T}{J}$ where $J$ is the coupling strength; a low effective temperature phase correpond to high values of $J$. For ReLU RBM, the probability distribution of configurations at temperature $T$ is defined as :

\begin{equation}
P_{{\bf w}}[{\bf v},{\bf h}]= e^{-\frac{E[{\bf v},{\bf h}]}{T}} \quad \text{with}\quad 
\frac{E[{\bf v},{\bf h}]}{T} = -\sum_i \frac{g_i}{T} v_i + \sum_\mu \left( \frac{h_\mu^2}{2 T} + \frac{h_\mu \theta_\mu}{T} \right) - \sum_{i,\mu} \frac{w_{i \mu}}{T} v_i h_\mu\ .
\end{equation}

Let ${\bf \bar{h}} = \frac{{\bf h}}{\sqrt{T}}$. The probability can be rewritten as $P[{\bf v}, {\bf \bar{h}}]=e^{-\bar{E}[{\bf v},{\bf \bar{h}}]}$
with 
\begin{equation}
\bar{E}({\bf v},{\bf \bar{h}}) = -\sum_i \frac{g_i}{T} v_i + \sum_\mu \left( \frac{\bar{h}_\mu^2}{2} + \bar{h}_\mu \frac{\theta_\mu}{\sqrt{T}} \right) - \sum_{i,\mu} \frac{w_{i \mu}}{\sqrt{T}} v_i \bar{h}_\mu\ .
\end{equation}
Since the marginal $P[\bf v]$ is not affected by the change of variable, a ReLU RBM at temperature $T$ is therefore equivalent to another ReLU RBM at temperature $T=1$, with new fields, thresholds and weights : $\bar{\bf{g}} = \frac{\bf{g}}{T}$, $\bar{\bf{\theta}} = \frac{\bf{\theta}}{\sqrt{T}}$, $\bar{\bf{w}} = \frac{\bf{w}}{\sqrt{T}}$. Therefore, changing the temperature is equivalent to rescaling the parameters, and in turn, the effective temperature of a given RBM can be deduced from the amplitude of its weights. For a R-RBM at temperature $T$:
$$ W_2 = \frac{1}{M} \sum_{\mu,i} \bar{w}_{i\mu}^2 \underset{N \rightarrow \infty}\sim \frac{p}{T} \ .$$
We therefore estimate the temperature of a given RBM  through $$ \hat{T} = \frac{\hat{p}}{\frac{1}{M} \sum_{\mu,i} w_{i\mu}^2}\ .$$ From this definition, it can be seen that the low temperature regime of the compositional regime, ${T} \ll {p} $, is equivalent to $W_2 \gg 1$. In RBM trained on MNIST, we typically find $W_2 \sim 7$

\subsection{Fields $g$}
Similarly to the weights, the fields $g_i$ and normalized fields could be estimated respectively as:
\begin{equation}
\begin{split}
\hat{g_i} &= \hat{T} \bar{g}_i \\
\hat{\tilde{g_i}} &= \frac{\hat{T}}{\hat{p}} \bar{g}_i = \frac{\bar{g}_i}{\frac{1}{M} \sum_{\mu,i} w_{i\mu}^2}
\end{split}
\end{equation}

A naive estimate for the normalized field $\tilde{g}$ would be to average the fields: $\hat{\tilde{g}} = \frac{1}{N} \sum_i \hat{\tilde{g_i}}$. It is however not really meaningful, as the $\hat{\tilde{g_i}}$ are extremely heterogeneous: for instance, the mean value over the sites $i$ of a single RBM is equal to $-0.48$, and is comparable to the standard deviation, $0.40$. This range of variation spans all the phases of R-RBM. To achieve quantitative predictions, we instead adjust the R-RBM parameter $g$ so that $q$, the mean value of $v_i$ in the visible layer, averaged over thermal fluctuations and quenched disorder, matches the value $0.132$ obtained from MNIST data. For the plots of Figure 4 in the main text, this gives $\frac{\hat{g}}{\hat{p}} = -0.1725$ for homogeneous R-RBM, and $\frac{\hat{g}}{\hat{p}} = -0.21$ for heterogeneous R-RBM.
\subsection{Thresholds $\theta$}
The thresholds and normalized thresholds can be estimated as 
\begin{equation}
\begin{split}
\hat{\theta}_\mu &= \sqrt{\hat{T}} \, \bar{\theta}_\mu \\
\hat{\tilde{\theta}}_\mu &= \sqrt{\frac{ \hat{T}}{\hat{p}}} \, \bar{\theta}_\mu  = \frac{\bar{\theta}_\mu}{\sqrt{\frac{1}{M} \sum_{\mu,i} w_{i\mu}^2}}
\end{split}
\end{equation}
Again, a naive estimate for the normalized threshold $\tilde{\theta}$ would be the average $\hat{\tilde{\theta}} = \frac{1}{M} \sum_\mu \hat{\tilde{\theta}}_\mu$ but this estimate is not meaningful. Indeed, contrary to the R-RBM case, the inputs $I_\mu$ of the hidden units $\mu$ are not evenly distributed around zero: $\mathbb{E} \left[ I_\mu \right] \neq 0$. Hence, even if the threshold is equal to zero, the activation probability can be different from 0.5. We take this effect into account by substracting the average value of the inputs from the average of $\theta$, and find that the difference is equal to $0.33$, with standard deviation $1.11$. This range of value for $\theta$ again spans all phases. In order to use a well-defined value, we choose $\theta$ such that the critical capacity $\alpha_c^{R-RBM}(\ell_{max}) = 0.5$, where $\ell_{max} \sim 1.5 $ is the maximum average index number observed across all RBMs trained for Fig. 4 in the main text. This estimation gives $ \hat{\tilde{\theta}} \sim 1.5$.